# VEHICULAR AD HOC AND SENSOR NETWORKS; PRINCIPLES AND CHALLENGES

Mohammad Jalil Piran[1], G. Rama Murthy[2], G. Praveen Babu[3]

[1]Department of Computer Science & Engineering, JNT University, Hyderabad, India
piran.mj@gmail.com
[2]Communication Research Centre, IIIT Hyderabad, India
rammurthy@iiit.ac.in
[3]School of Information Technology, JNT University, Hyderabad, India
pravbob@jntu.ac.in

## ABSTRACT

*The rapid increase of vehicular traffic and congestion on the highways began hampering the safe and efficient movement of traffic. Consequently, year by year, we see the ascending rate of car accidents and casualties in most of the countries. Therefore, exploiting the new technologies, e.g. wireless sensor networks, is required as a solution of reduction of these saddening and reprehensible statistics. This has motivated us to propose a novel and comprehensive system to utilize Wireless Sensor Networks for vehicular networks. We coin the vehicular network employing wireless Sensor networks as Vehicular Ad Hoc and Sensor Network, or VASNET in short. The proposed VASNET is particularly for highway traffic .VASNET is a self-organizing Ad Hoc and sensor network comprised of a large number of sensor nodes. In VASNET there are two kinds of sensor nodes, some are embedded on the vehicles-vehicular nodes- and others are deployed in predetermined distances besides the highway road, known as Road Side Sensor nodes (RSS). The vehicular nodes are used to sense the velocity of the vehicle for instance. We can have some Base Stations (BS) such as Police Traffic Station, Firefighting Group and Rescue Team. The base stations may be stationary or mobile. VASNET provides capability of wireless communication between vehicular nodes and stationary nodes, to increase safety and comfort for vehicles on the highway roads. In this paper we explain main fundamentals and challenges of VASNET.*

## 1. INTRODUCTION

As reported by the National Highway Traffic Safety Administration (NHTSA), the car accidents on highways have an ascending rate. In the US only, vehicles' crashes on the highways resulted in the loss of as many as 40,000 lives and overall economic losses of more than $230 billion. These lugubrious statistics have motivated both academician and industrial researches to find out a solution. However, Vehicular networks have received intensive of research work in the recent years

due to the wide variety of services they provide. Vehicular Ad Hoc Network (VANET) is the most important component of Intelligent Transportation System (ITS) [1], in which vehicles are equipped with some short-range and medium-range wireless communication. In VANET two kinds of communication are supposed: Vehicle-to-Vehicle and Vehicle-to-road side units, where the road side units might be cellular base station for example. From the definition of VANET, a salient challenge is obvious. Suppose at the mid-night in some rural area, a vehicle has a very important data packet (i.e. detection of an accident) which should be forwarded to the following vehicles immediately. The probability of low density of vehicles in the rural areas at mid-night is very high. Consequently, in this situation the packet will be lost due to lack of presence of other vehicles to receive and broadcast it, and arrival of the following vehicles in the accident area is unavoidable. If the above discussed accident occurs in a tunnel and fire takes place, a tragedy may turned out by presence of the other vehicles; do remember tragedy of tunnel of the Mont-Blanc between France and Italy [5]. To overcome this serious issue, we suppose to utilize wireless sensor nodes on both sides of the highway too. We call our proposed system as Vehicular Ad Hoc and Sensor Networks or in short VASNET.  The motive behind VASNET is safety on highway roads, since many lives were lost and many more injuries have been occurred because of the car accidents. There are two types of sensor nodes in suggested VASNET, some are embedded in the vehicles – known as vehicular Nodes (VN)- and others are deployed in predetermined distances besides the highway road, known as Road Side Sensor nodes (RSS). We can have some Base Stations (BS) such as Police Traffic Station, Firefighting Group and Rescue Team. The base stations may be either stationary or mobile. The VNs are supposed to collect the real data such as vehicle' velocity, and forward towards BSs via RSS nodes. On the other hand, for sending a query from BSs, RSS are supposed nodes receive it and rebroadcast the packet to the vehicles in its coverage area.

The rest of this paper is organized as follows. First, section 2 describes the related works. This is followed by explanation about VASNET Topology in section 3. Section 4 addresses VASNET Communication Architecture. Conceivable applications are suggested in section 5. Some necessary future works are mentioned in 6. Finally, section 7 concludes the paper.

## 2. RELATED WORKS

### 2.1 GPS-based Vehicular Networks

Vehicles gain benefit of Global Positioning System (GPS) in, an Offline and Online Mode too. A device - GPS receiver-embedded on vehicles was deemed in both modes, which can receive the satellites' signals and estimate the position of the vehicle. An MMC Card is required for each vehicle to save data in an offline mode, whereas in the online mode, a GMS is exploited to send data to the station by the SMS format. The data stored in the MMC Card can be retrieved via sophisticated software. For the online mode, an industrial mobile hardware is used for data interpretation [2].

Disadvantages:

a) GPS's signals are under the effect of the following which attenuate them: [3]

1. Delay of Troposphere (the lowest portion of Atmosphere) and Ionosphere: Satellites signals become weak when they pass through the atmosphere.
2. Multiple Signals; occurs when GPS signals reflect by the buildings or rocks before reaching to the receiver.
3. Receiver Periodical Errors: Surely receiver's time is not working as proper as GPS satellites; therefore it is prone to high errors about time meters.
4. Orbit Error: Temporary data might not report the exact location of the satellite.
5. Obstacles: Some other satellites, buildings, trains, electronic obstacles, crowded trees can prevent signals.
6. Satellites Geometry: Satellites geometry is pointed to the proportional location of satellites. When the satellites are on the same way or they are in the small groups, some geometry errors happen.
7. Satellite's signal intentional corruption: This was made by Defence Organization to prevent using of robust signals of GPS satellites by unauthorized people.

b) Hardware constraints:

1. The necessity of additional hardware as GPS receivers, MMC Card, SIM CARD, ...
2. Less accuracy (up to 15 meters in positioning and 0.5 km/h for velocity).
3. Dependency on GPRS system in online mode.
4. Failure of MMC Card in the offline mode.

Therefore, GPS cannot be a good solution.

## 2.2 Vehicular Ad Hoc Network (VANET)

Vehicular Ad Hoc Networks (VANET) upon implementation should collect and distribute safety information to massively reduce the number of accidents by warning drivers about the danger before they actually face it [4]. VANET consist of some sensors embedded on the vehicles. The onboard sensors' readings can be displayed to the drivers via monitors to be aware of the vehicle condition or emergency alarms, and also can be broadcasted to the other adjacent vehicles. VANET can also be helped by some of Roadside Units like Cellular Base Stations, to distribute the data to the other vehicles. VANET makes extensive utilization of wireless communication to achieve its aims. VANET is a kind of Mobile Ad Hoc Networks (MANET) with some differences, like (1) Limitation in Power: in MANET, power constraint is one of the most important challenges which has shadowed all other aspects namely routing, fusion, on the other hand in VANET, huge battery is carried by the vehicle (i.e. car's battery), so, energy consumption is not a salient issue, (2) Moving pattern: which is random in the MANET while vehicles tend to move in an organized fashion in VANET, and (3) Mobility: there is high mobility in the VANET in comparison to MANET. However, self-organization and lack of infrastructure are similarities between MANET and VANET. There are some salient challenges in VANET such as; (1) as mobile nodes (vehicles) moving with high mobility, therefore quick changes in the VANETs topology are difficult to control. (2) The communication between the vehicles is prone to frequent fragmentation. (3) Rapid change in link's connectivity cause many paths to be disconnected before they can be utilized. (4) There is no constant density in VANET, as in highways high density and in the rural low there is density. (5) A message can change the topology, for instance, when a driver receives an alarm message, s/he may changes his/her direction, which may cause the change the topology.

## 3. VASNET TOPOLOGY

 VASNET inherits its characteristics from both Wireless Sensor Networks (WSN) and Vehicular Ad Hoc Networks (VANET).  There is no infrastructure for VANET, therefore the vehicular nodes do perform data collection as well as data routing.  Therefore, the necessity of designing a new architecture to overcome the mentioned challenges is transpicuous. In this paper, we propose a novel topology, which can be a suitable solution to over come VANET issues.

VASNET is a fusion of WSNs and MANET, which can be divided in to three layers. The upper layer consisting of traffic monitor stations, e.g. traffic police located at the cities. These are connected by either fiber optic cables to form the backbone of traffic information network. The middle layer is region layer, consisting of traffic check post located through highways. These stations can be connected via the Internet or local networks, and finally the lower layer is the field layer, consisting of WSN nodes deployed on beside the highway and onboard sensors which are carried by the vehicles. These nodes are connected by short-range or medium-range wireless communication. The components are as follows:

    (1) Vehicular Sensor Nodes; which are carried by the vehicles. These nodes are supposed to sense the real phenomena e.g. the velocity of the vehicle. The sensor readings are to be sent to the base stations via RSS nodes. These nodes can communicate with each other or the roadside sensor via short-range communication.

    (2) Road Side Sensors (RSS); are deployed in a fixed distance beside the road. RSSs act as cluster heads for vehicular nodes. RSS nodes receive the data from mobile nodes and retransmit towards the BSs. These nodes are equipped with two kinds of antenna, unidirectional and bidirectional. Unidirectional antenna is for broadcasting and directional antenna are intended for geo-casting. We need to satisfy the following requirements for deploying the sensor nodes on a road side, such as; (a) high reliability, (b) long time service and (c) high real time.

    (3) Base Station (BS); are Police Traffic Control Check-Post, Rescue Team Buildings or Fire Fighting Stations in some fixed point trough the roads. We can have mobile BS like, Traffic Police patrolling team, Firefighting Truck, or ambulance.

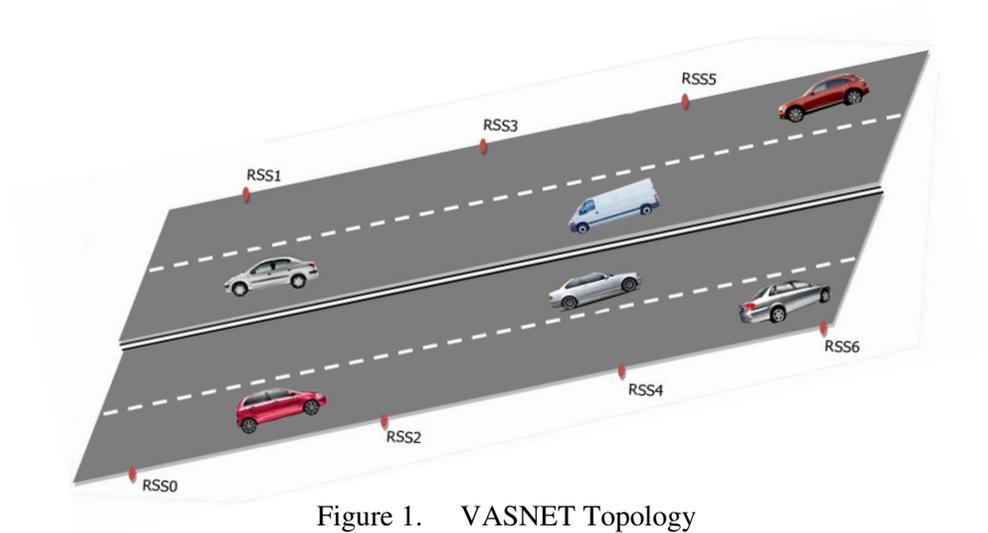

Figure 1.    VASNET Topology

## 4.  VASNET COMMUNICATION ARCHITECTURE

The sensor nodes are deployed in vehicles as well as both sides of highway roads. Vehicular' nodes have the ability to collect imperative data and route data to the base stations. Sensor readings are routed to the end users by multi-hop infrastructure architecture via intermediate nodes i.e. RSS nodes. The protocols stack which may be used by CR-VASNET nodes is given in figure 2. The protocol stack consists of five layers and three planes. The planes are to help sensor node o coordinate the sensing tasks and lower overall power consumption. More specifically, the power management plane, manages power consumption for example defining sleep and wake status for the nodes. The mobility management plane monitors the movement of sensor nodes, so a route back to the user is always maintained. And finally, the task manager plane balance and coordinates the sensing tasks given to a particular given region. In the following subsections we investigate and briefly explain the specific design consideration of each communication layer of nodes with respect to dynamic spectrum management.

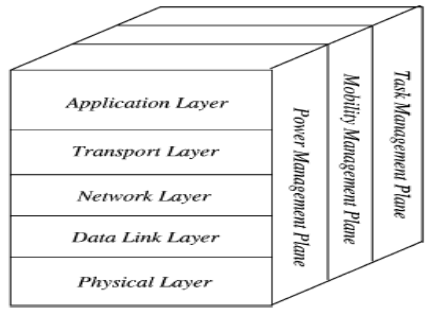

Figure 2.    The Protocol Stack [1]

### 4.1. Physical Layer

Federal Communications Commission (FCC) [17] assigned a new 75 MHz band Dedicated Short Range Communication (DSRC) at the 5.9 GHz frequency for Intelligent Transportation Systems (ITS) applications in North America. The band is divided into seven channels as abstracted in the following figure 3 [18].

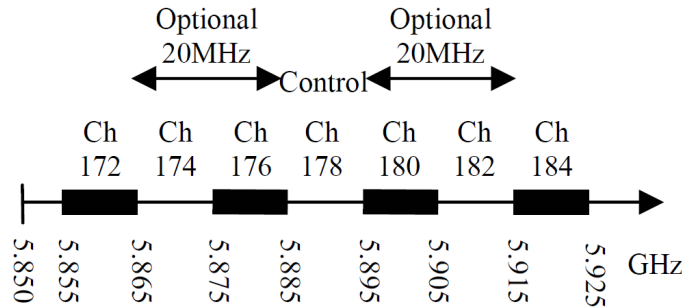

Figure 3.    DSRC Bands

A physical layer standard is being developed by the American Society for Testing and Material (ASTM), known as the ASTM E2213 standard. It uses Orthogonal Frequency Division Multiplexing (OFDM) as its modulation scheme and covers distances up to 1 km [19]. Since OFDM is multi carrier modulation, data is split into multiple lower rate streams and each stream us used to modulate one of the subcarriers. Due to reduction of data rate, lower bandwidth is necessary for each carrier. Although high data rates can be achieved using OFDM. The performance of OFDM can degrade rapidly if careful considerations for synchronization and channel variations are not taken. OFDM is sensitive to frequency and phase errors [20, 21]. In VASNET the high mobility of the vehicles on opposite sides or between the vehicles and RSS nodes cause an increase in the received frequency. This important subject must be considered during the design of he receiver as it destroys the orthogonality of the carriers and increases ICI [22]. The 48 carriers out of 64 carriers assigned by IEEE 802.11 are utilized for data, four of them are pilot carriers and the other are not used to reduce interference to other bands. Training sequences are used at the beginning of the packet for training and the pilot carriers channel response is extrapolated to estimate the channel response for the other carriers [23].

### 4.2. Data Link Layer

Generally, this layer is responsible for reliable sending and receiving of data frames between communicating nodes. The most important functionality of data link layer is medium access control (MAC). In VASNET, event driven messages should have the higher priority compared to other messages, like periodic and comfort message. Therefore, some mechanisms for service differentiation and admission control are indispensable. In fact, we can define three levels of priority for messages in VASNET: (1) event driven safety messages, (2) beaconing safety messages, and (3) comfort messages respectively in descending order. The required mechanisms are dependent on MAC layer policy. Recently IEEE 802.11a was selected by American Society for Testing and Material (ASTM) to be the fundamental for its standards of Dedicated Short Range Communication and IEEE 1609 group proposed DSRC [16] as IEEE 802.11p standard [12]. It is a promising MAC mechanism for VASNET. Furthermore, MAC layer based on UTRA TDD [13] enhanced by CarTALK can be an acceptable and practical mechanism for VASNET.

### 4.3. Network Layer

VASNET as an application of WSN needs special multi-hop wireless routing algorithm between the sensor nodes –both mobile and stationary- and the base stations. VASNET inherits network layer issues from traditional wireless sensor networks [3], [6], and mobile Ad Hoc networks (MAENT) [7], [8], [9] such as infrastructure-less, unstable topology, multi-hop networking, energy efficiency data-centric routing, attribute-based routing, localization, etc.

Proactive routing protocols are not feasible for dynamic topologies. Therefore, it seems that reactive routing algorithms are more suitable. However, the communication overhead and increased contention, dynamic spectrum aware reactive routing can be considered for VASNET.

Since VNs can receive significant power from the car's battery, they are not restricted in energy. On the other hand, RSS nodes have no such power sources, so energy must be consumed efficiently. It is desirable to put the nodes in a low power sleep mode to minimize average energy consumption. In [2] we proposed an energy efficiency routing algorithm for VASNET, in which we kept some of RSS nodes in sleep mode to save power. The decision whether to stay awake or in sleep mode is decided by a probability which is decided by (1) Remaining energy in the sensor, (2) Generated random number, (3) Previous sate of the node and (4) Importance of the message. The following figures show the obtained results by simulation of the proposed routing algorithm.

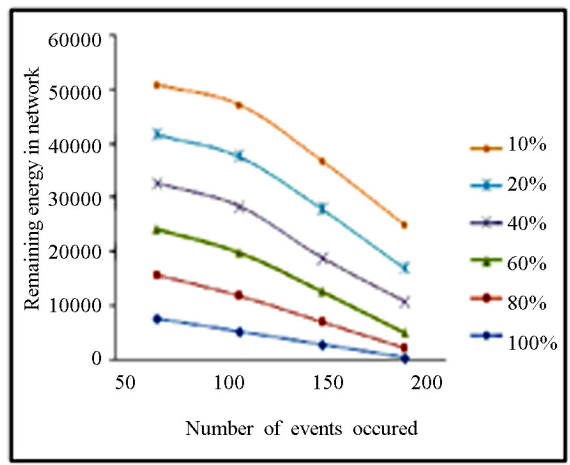

Figure 4.    Remaining Energy VS Number of events.

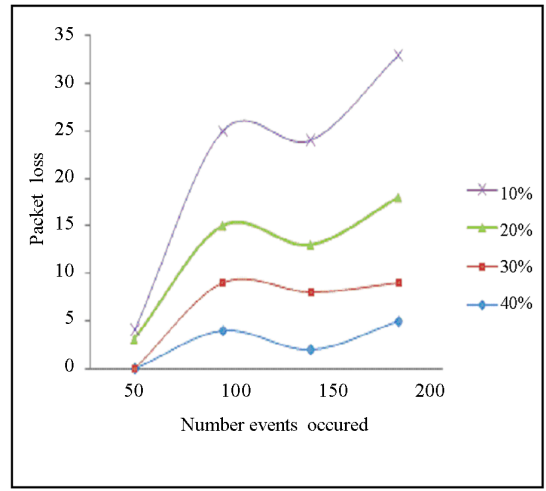

Figure 5.    Packet Loss VS Number of events.

### 4.4. Transport Layer.

End-to-End reliable delivery, congestion control to preserve scarce network resources, e.g. energy, and taking care of application based QoS requirements are the major functionalities of transport layer. When sensor nodes detect an event, they try to send their readings towards base station. This makes a bursty traffic in to the network. Significant sensory data must be reliably delivered to the base station to obtain detection and tracking an event signal. Simultaneously, if the multi-hop network capacity exceed, congestion is the result. However, congestion is one of the obstacles to achieve to energy efficiency. There is delicate balance between reliability and energy efficiency, which has been the main focus of transport layer solution proposed for wireless sensor networks so far [10]. While the mentioned balance between reliability and energy efficiency is owned by VASNET, designing an operative transport layer is an open research issue.

### 4.5. Application Layer

In case of wireless sensor networks, application layer may be responsible for some functions like generation of information, interest and data dissemination, feature extraction of event signals, methods to query sensors, and data aggregation and fusion. However, in our proposed system, these services have to utilize the capabilities of VASNT while conforming to its constraints. Therefore, designing of new protocols for this layer with respect to its pros and cons is required.

## 5. CONCEIVABLE APPLICATIONS FOR VASNET

VASNET will play an important role for future automotive development sine it has capability of providing wide range of service from safety, comfort, telematics and infotainment to diagnosis being the primary design requirements. Although sensors installed on vehicles are limited to have others sensors in their communication range, existence of roadside sensors extend the coverage range to multiple kilometres. In VASNET, vehicles and roadside sensors complement the local sensor data with sensor data received from the other vehicles [11]. Furthermore, the specific properties of the system allow the deployment of a wide range of new attractive application and services which are neither feasible nor cost-effective with the other systems such as VANET.

VASNET presents a wide range of applications in the subjects of safety, convenience. Safety applications monitor the vehicles' velocity, distance between vehicles, surface of the road, etc. Convenience applications need to enhance traffic efficiency by boosting the degree of convenience for drivers and commercial applications will provide some entertainment and service such as web access and streaming audio and video streaming.

Following are some of the conceivable applications of the VASNET:

### 5.1. Safety Application:

- Velocity Monitoring; the velocity is under control of the vehicular sensor nodes. When the vehicle's speed is more than the authorized velocity, a report will be sent to the station via RSS nodes.

- Accident Notification, a vehicle can estimate the accident, if the accident takes place. Then with respect to existence of RSS nodes, it can calculate its position, and finally it sends the packet to the BSs e.g. Traffic Police, Rescue Team, Fire Fighting Groups, highway patrol for tow away support and also to the trailing vehicles.

- Road Hazard Control Notification; deals with cars notifying other cars about land sliding or existence of any obstacles such as rocks or snow.

- Cooperative Collision Warning; warns the drivers potentially under crash route so that they can mend their ways.

- Cooperative collision avoidance.

### 5.2. Convenience Application:

- Theft Report; for a stolen vehicle, after the owner's report to the police, a query will be sent from the traffic station to all RSS nodes to find out the vehicle. The receiver of the packet then, broadcast it to the vehicles along road, the wanted vehicle's sensor will prepare a reply to the RSS, and the RSS adds the location and send back to the station.

- Electronic Penalty Bills; when a vehicle's sensor node detect an offence, such as high velocity, it sends a report to the traffic station and after its confirmation, the sensor will save the record in its database, e.g. date, time, location, etc. this data can also be updated through online.

- Congested Road Notification; detects and notifies about road congestions which can be used for routing and trip planning.

- Traffic Information for dynamic route updates, depending on existing obstruction of traffic e.g. by constriction or traffic jams.

## 6. FUTURE WORKS

There are a lot of unexplored topics which need plenty of research work in VASNET, such as (1) data fusion to reduce the number of transmissions and subsequently save the energy, (2) localization, to find out the location of a vehicle on the highway, (3) spectrum access, due to spectrum scarcity, (4) security and many more other aspects. In this section, we describe data fusion and localization:

### 6.1. Data Fusion

Diverse sensors today are used and the data received is integrated to provide a unified interpretation, which has become an area of research interest. The best and most considerable advantage of sensor fusion is, to obtain high level information in both statistical and definitive aspects, which can not be attained by a single sensor. In the other words, data fusion is a process of combining information from several sources to educe beneficiary and reliable information. Data Fusion is supposed to perform: (1) detection of existence of an object, an event or a process, (2) identification of an object's status, (3) event monitoring, (4) physical relation with an object, or (5) independent information relation to take an intelligent decision. In VASNET, accident intensity evaluation can be done via data fusion using fuzzy theory. Because fuzzy logic [14] [15] methods have the ability of fusing uncertain data from several sources, in order to improve the quality of information.

### 6.2. Localization

When an accident takes place on a highway, warning messages must be sent to the BSs, e.g. Traffic Police, firefighting and rescue team, as well as to inform all the following vehicles. Computing the location of vehicles, in case of accident or high velocity (more than authorized speed), is very essential for the safety of vehicles. Although GPS was introduced as a good solution, but as we described before, it is not available anywhere and anytime, particularly in surrounding areas like tunnels. Localization for vehicular networks has becomes an interesting research area in the recent years. However, there are a lot of proposals for calculating the position of a vehicle such as: (1) Track Detection (TRADE), (2) Distance Defer Time (DDT), (3) Optimize Dissemination of Alarm Messages (ODAM), (4) Role Based Multicast (RBM), etc. Since RSS nodes are aware about their location, a vehicular node can calculate its position with respect to RSS nodes' location. The vehicular node may ask the RSS nodes in its coverage range to send their location coordinates. Then the VN, after receiving at least three replies from RSS nodes will be able to calculate its position by triangular formulas.

## 7. CONCLUSION

This paper presented a state-of-the-art survey in Vehicular Ad hoc and Sensor Network (VASNET), a fusion of Wireless Sensor Networks (WSN) and Mobile Ad Hoc Networks (MANETs), as a promising approach for future intelligent transportation system (ITS). However, we introduced some aspect and involved challenges. We explained feasible topology and communication architecture applicable to VASNET. And also conceivable applications were introduced. VASNET as a novel scenario for vehicular networks need a lot of research works. Hence, we state some of its requirements e.g. data fusion, localization and spectrum access.